\documentclass[%
reprint,
superscriptaddress,
bibnotes,
 amsmath,amssymb,
 aps,
prc,
preprintnumbers
]{revtex4-1}

\usepackage{graphicx}
\usepackage{dcolumn}
\usepackage{bm}

\newcommand{\ket}[1]{\left| #1 \right \rangle}

\usepackage{bm, color}
\usepackage{lipsum}

\newcommand{\mfr}{\mathbf{r}}

\newcommand{\up}{\uparrow}
\newcommand{\dn}{\downarrow}

\newcommand{\ac}{\tilde{a}}
\newcommand{\rec}{\tilde{r}_{\textrm{e}}}

\definecolor{ryangreen}{rgb}{0.20,0.8,0.0}

\definecolor{red}{rgb}{0.81,0.13,0.16}

\usepackage{hyperref}
\hypersetup{
    colorlinks=true,
    linkcolor=blue,
    filecolor=magenta,      
    urlcolor=blue,
    citecolor=blue,
    }

\begin{document} 

\preprint{LA-UR-25-31665}

\title{Exploring the shell structure of trapped superfluid gases}

\author{Georgios Palkanoglou}
\affiliation{TRIUMF, 4004 Wesbrook Mall, Vancouver, BC V6T 2A3, Canada}
\author{Ryan Curry}
\affiliation{Theoretical Division, Los Alamos National Laboratory, Los Alamos, NM 87545, USA}
\affiliation{Department of Physics, University of Guelph, 
Guelph, ON N1G 2W1, Canada}
\author{Stefano Gandolfi}
\affiliation{Theoretical Division, Los Alamos National Laboratory, Los Alamos, NM 87545, USA}

\begin{abstract} 
We provide quantum Monte Carlo calculations of a two-component Fermi system interacting with
an attractive interaction confined in harmonic traps. We investigate the role of the interaction's
scattering length and effective range and show its important role in modifying shell effects in the
structure of these systems. We show that in the strongly interacting regime, where the scattering
length is large, the dominant role to shell effects is due to the effective range. These conclusions
are very relevant for nuclear physics, in particular for neutron-rich systems, and open the way to
perform new atomic experiments that can help to explain the disappearance of shell-effects in nuclei.
\end{abstract}

\maketitle 
\section{Introduction}
Cold atomic Fermi gases have received considerable attention in the past two decades on the experimental and theoretical fronts alike~\cite{Ketterle_Zwierlein_2008}. For homogeneous strongly interacting Fermi gasses, several properties have been measured and calculated~\cite{Giorgini_Pitaevskii_Stringari_2008,Randeria_Taylor_2014}, including the equation of state~\cite{Ku_Sommer_Cheuk_etal_2012,Navon_Nascimbene_Chevy_etal_2010}, the spin and density response at high momentum transfer~\cite{Hoinka_Lingham_Delehaye_etal_2012}, and properties related to contact parameters~\cite{Hoinka_Lingham_Fenech_etal_2013}. 
Inhomogeneous gases have received similar attention in
studies of the three- to two-dimensional transition~\cite{Sommer_Cheuk_Ku_etal_2012},
fermions in optical lattices~\cite{Bloch_Dalibard_Zwerger_2008,Ma_Pilati_Troyer_etal_2012,
Pilati_Zintchenko_Troyer_2014}, and small clusters of cold fermions at
unitarity~\cite{Zurn_Wenz_Murmann_etal_2013,Wenz_Zurn_Murmann_etal_2013}. In the latter case it has become possible
to confine and experimentally control very few atoms
in a trap~\cite{Murmann_Deuretzbacher_Zurn_etal_2015}. When driven to strongly interacting regimes, these experiments can provide insight into the structure of more complicated strongly interacting finite systems, like the atomic nuclei where the strong nuclear force creates self-bound systems of neutrons and protons. 

A fundamental property of finite fermionic systems, strongly interacting or otherwise, is their shell structure which manifests as abrupt changes in the system's microscopic structure with the addition of particles. While initially identified in electronic systems, where electrons are arranged in atomic orbitals~\cite{Pauli_1925}, these effects are ubiquitous in atomic nuclei where the presence of ``magic numbers'' of neutrons or protons plays a central role in the description of nuclear structure~\cite{Sorlin_Porquet_2008,Janssens:2009aa,Wienholtz:2013nyaa,Steppenbeck:2013mga,Ding:2025hpq}. These configurations have been experimentally observed to correspond to significantly higher binding (or reduced cross-sections), similar to the higher ionization energy
required to excite an atom whose electrons form a closed-shell  configuration, and explained in a mean-field picture as arising from an average nuclear potential. Similar effects are anticipated in few-atom systems in a trap where one can look for universal patterns translating to a more complicated finite system like the atomic nucleus. Investigating cold-atomic systems for such universal properties that are transferable to nuclear systems is a common interface of the two fields that is underpinned by the proximity of the unitary Fermi gas to nucleonic matter~\cite{Strinati:2018wdg,Ohashi:2020djc}. This interface has been exploited before to understand the properties of low-density neutron matter, found in the inner crust and outer core of neutron stars or the exterior of neutron-rich nuclei~\cite{Gandolfi:2015jma}. In this paper we will explore the latter association in understanding some exotic concepts in nuclear structure, namely the shell structure of neutron-rich systems.

Magic numbers have been observed to disappear~\cite{Liu_Obertelli_Doornenbal_etal_2019} or shift~\cite{Janssens:2009aa,Wienholtz:2013nyaa,Steppenbeck:2013mga} in exotic and neutron-rich nuclei. A half century after first identified~\cite{Thibault:1975zz}, this concept is now named ``shell evolution'' and it still attracts considerable attention from nuclear theory and experiment alike~\cite{Otsuka:2018bqq,Hagen:2012sh,Steppenbeck:2013mga}. Shell evolution lies on the forefront of current research in exotic nuclei, in many cases being understood to originate from various intricacies of the nuclear many body problem, such as certain symmetries of the nucleon-nucleon interaction~\cite{Otsuka:2001nw, Ding:2025hpq}, many nucleon forces~\cite{Stanoiu:2004nm,Kanungo:2009zz}, exotic many-body phenomena such as neutron halos~\cite{Otsuka:2018bqq}, and so on; for a review on the various identified mechanisms contributing to shell evolution see Ref.~\cite{Otsuka:2018bqq}. However, many fully or partially unresolved questions remain, especially in heavy nuclei and close to the drip-lines~\cite{Nowacki:2021fjw,Otsuka:2018bqq}. As the nuclear experiment pushes to heavier and more weakly bound systems, such topics are expected to become increasingly relevant~\cite{Mollaebrahimi:2025aa,Brown:2024rml}.

Another fundamental property of fermionic systems is their tendency to form pairs which is the mechanism driving fermionic superfluidity and superconductivity. Neutrons are known to give rise to superfluid states as a condensation of pairs in spin-singlet states at low densities~\cite{Dean:2002zx,Gandolfi:2022dlx}, pairs in spin-triplet states in higher densities ~\cite{Ding:2016oxp}, or even more exotic structures~\cite{Ma:2026dti}. The corresponding superfluid effects induced by the presence of neutron-neutron pairing correlations are essential for the description of neutron-rich matter and can be found in the rotational and thermal properties of neutron stars~\cite{Sedrakian:2018ydt,Watanabe:2017nzj,Page:2004fy} as well as most properties of finite nuclei, most notably the increased binding and compactness of even-mass nuclei~\cite{Dean:2002zx}. In finite systems pairing correlations are known to antagonize shell-structure, with pronounced shell-closures weakening the pairing correlations, which typically arise mid-shell, and with dominant pairing smearing the discontinuities induced by shell-closures~\cite{Dean:2002zx,Litvinova:2011vp}. In total, the interplay of pairing and shell-structure in nuclei often gives rise to unexpected effects~\cite{Dobaczewski:1995bf,Baroni:2009eh,Simpson:2026gwe}.

In this paper we report Quantum Monte Carlo (QMC) calculations of the ground state energy of a two-component Fermi gas (spin-up and down particles) confined in a harmonic oscillator (HO) well,
interacting with a two-body attractive interaction. QMC methods provide a nonperturbative solution to the many-body Schr\"{o}dinger equation, and are widely employed in condensed matter, cold atomic and nuclear physics \cite{Foulkes:2001zz, Carlson:2012mh, Carlson:2014vla}. In particular, we employ the diffusion Monte Carlo (DMC) method to perform \textit{ab initio} calculations of the ground state energy of even and odd atom numbers. 

We explore the shell structure of the two-component Fermi gas by calculating two-particle separation energies and two-particle shell gaps, and we make connections to its pairing correlations via calculations of the superfluid pairing gap. We explore these connections at and approaching unitarity in search for universal properties applicable to the shell-structure of low-density neutron matter like the one found in the exterior of heavy neutron-rich nuclei. The inherent interest in shell-structure lies in its utility as an average property underlying the structure of many nuclei and as such one can expect that from a first principles approach its details are a compound effect of the properties of the forces that bind the nucleus and their interplay with many-body physics. In this paper we provide evidence for the pairing correlation's contribution to the disappearance of magic numbers in neutron rich nuclei as that originates from the proximity of low-density neutron matter to the unitary Fermi gas.

\section{Hamiltonian}
Unitarity is typically identified as the limit where a zero-range two-body interaction is strong enough to almost create a free-space bound state. At that limit the physics of the system become independent of the details of the two-body potential. Thus unitary fermions in a harmonic trap can be described with a many-body Hamiltonian, 
\begin{align}
    H&=-\frac{\hbar^2}{2m}\sum_{i=1}^N\left(\boldsymbol{\nabla}^2_i+\frac{1}{2}m\omega^2\mathbf{r}_i^2\right) + \sum_{i<j'} v(r_{ij'}) \label{eq:hamiltonian}
\end{align}
where the two-body interaction is a s-wave short-range potential where $\sum_{i<j'}$ runs only over pairs of opposite-spin particles.
We use a modified P{\"o}schl-Teller (PT) potential of the form,
\begin{align}
v(r) = -v_0 \frac{\hbar^2}{2m}\frac{\beta^2}{\cosh^2(\beta r)},
\end{align}
with parameters $v_0$ and $\beta$ that can be tuned such that the two-body interaction reproduces certain scattering length $a$ and effective range $r_e$. This tunable quality makes the PT potential very useful for QMC studies of strongly interacting systems in both cold atomic and nuclear physics \cite{Carlson_Chang_Pandharipande_etal_2003, Gezerlis_Gandolfi_Schmidt_etal_2009, Gezerlis_Carlson_2010, Gandolfi_Schmidt_Carlson_2011, Forbes_Gandolfi_Gezerlis_2011, Madeira_Gandolfi_Schmidt_etal_2019, Madeira_Frederico_Gandolfi_etal_2021, Curry_Lynn_Schmidt_etal_2023, Gandolfi_Curry_Gezerlis_2024}. In this parametrization the unitarity regime corresponds to $v_0=1$ and $\beta/R \gg 1$, where $R$ is the interparticle distance. Specifically, the form of the PT potential allows for a straightforward identification of the effective range since for $v_0=1$: $r_e(1,\beta) = 2/\beta$. Again, the functional form of the potential in Eq.~(\ref{eq:hamiltonian}) is not important: our purpose is the study of shell effects close to the unitarity regime where the details of the interparticle interaction become irrelevant.

The harmonic oscillator potential introduces a length scale $R_0=\sqrt{\hbar/m\omega}$ which can be used to replace the interparticle separation, in lieu of a constant density inside the trap: $R_0^{-3}$ can be used as a scale for the particle density defined by the confining potential. One can then define unitarity as $\ac=a/R_0 \gg 1$ and $\rec=r_{\textrm{e}}/R_0\ll 1$. As we will see in the next sections, the ground state properties of the trapped fermions depend strongly on $\ac$ and $\rec$. 

\section{Quantum Monte Carlo}
We solve for the ground state of the Hamiltonian in Eq.~(\ref{eq:hamiltonian}) using DMC which solves the many-body Schr{\"o}dinger equation by recasting it as a diffusion equation in imaginary time. Assuming a trial-wavefunction $\ket{\Psi_T}$ that is not orthogonal to the ground state of the many-body Hamiltonian, $\ket{\Psi_0}$, DMC extracts the ground state via imaginary-time evolution:
\begin{align}
    \lim _{\tau\to \infty} e^{-\tau(H-\hbar)/\hbar} \ket{\Psi_T} \propto \ket{\Phi_0}~.
\end{align}
In practice, the limit $\tau\to\infty$ is taken by distributing particles according to the trial wavefunction and propagating them stochastically through a repeated sampling of the short-time propagator. The speed of convergence depends largely on the choice of the trial state where the most efficient are wavefunctions that provide a good qualitative description of Hamiltonian's ground state.

In that spirit we use a trial wavefunction inspired by the BCS theory, the mean-field description of superfluids, as has been done often in the past:
\begin{align}
    \Psi_T &= \prod_{ij'}f_J(r_{ij'}) \Phi_{BCS} \label{eq:trial1}\\
    \Phi_{\textrm{BCS}}&=\mathcal{A}\left[\phi(\mfr_{1},\mfr_{1'})\phi(\mfr_{2},\mfr_{2'})\cdots \phi(\mfr_{n},\mfr_{n'})\right]~, \label{eq:trial2}
\end{align}
where the (un)primed indices signify $\up$($\dn$)-spin particles, running from 1 to $n=N/2$ and $\mathcal{A}$ is the antisymmetrization operator that ensures antisymmetry among like-spin particles. Due to the antisymmetrizer, $\Phi_{\textrm{BCS}}$ contains nodes and as such captures most of the structure of the trial state. In contrast, the symmetric (and nodeless) Jastrow factor $f_J(r_{ij})$ is employed for convenience as it reduces the variance of the stochastic propagation. This is standard practice and we follow the choice made in Ref.~\cite{Carlson_Gandolfi_vanKolck_etal_2017}. Finally, the pair wavefunctions $\phi$ are taken to be similar to those used in Ref.~\cite{Chang_Bertsch_2007} and take the form,
\begin{align}
    \phi(\mfr_i,\mfr_j) = \sum_{k=1}^{N_c} c_k \sum_{m=-l_k}^{l_k} \frac{(-1)^{l_k+m}}{\sqrt{2l_k+1}} \times \notag \\
    \times \varphi_{n_kl_km}(\alpha \mfr_i)\varphi^\star_{n_kl_km}(\alpha \mfr_j)~, \label{eq:pair_wf}
\end{align}
where $\varphi_{n_kl_km}$ is the single-particle harmonic oscillator orbital of shell $n$ and angular momentum ($z$-projection) $l$ ($m$). The $c_k$ and $\alpha$ are variational parameters, and we take $N_c=12$. This form of the pair wavefunction ensures a total angular momentum of zero. Compared to Ref.~\cite{Chang_Bertsch_2007}, the pairing wavefunction in Eq.~(\ref{eq:pair_wf}) contains the extra variational parameter $\alpha$ which we find essential in lowering the variational energy. We also tried to use the more general wavefunction of Ref.~\cite{Carlson_Gandolfi_2014} but found very similar results. All the variational parameters were optimized using the stochastic reconfiguration method~\cite{Sorella_2001} in the context of a variational Monte Carlo (VMC) preceding each DMC calculation (for details see Ref.~\cite{Carlson_Gandolfi_2014}).

\begin{figure}[t!]
\includegraphics[width=0.98\linewidth]{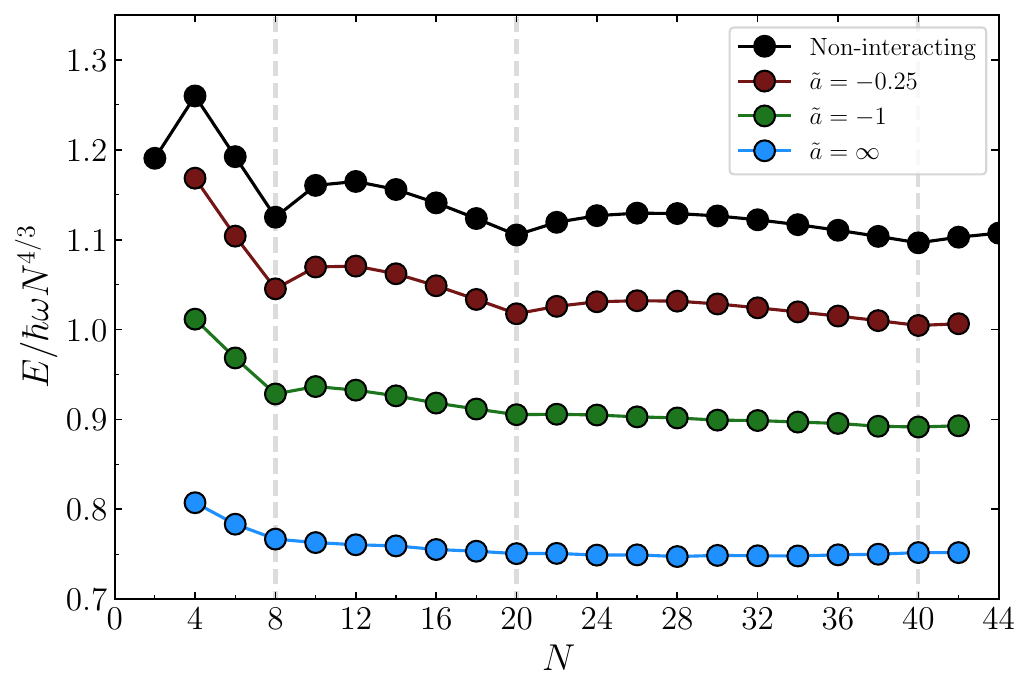}
\caption{The energy as a function of the number of particles for different 
scattering length and fixed effective range.}
\label{fig:as}
\end{figure}

\section{Trapped Fermions Approaching Unitarity}
\subsection{Scattering length}
We want to investigate the shell effects of $N$ unitary fermions and as such most of our discussion will revolve around that limit of large scattering length $a$ and small effective range $r_e$. We first investigate the ground state energy as a function of the scattering length $a$ which is tuned via the parameters $v_0$ and $\beta$ in Eq.~(\ref{eq:hamiltonian}). In Fig.~\ref{fig:as} we plot the ground state energy for $\ac=-0.25, -1, -\infty$ while keeping the effective range small, $\rec=0.1$, that is, approaching unitarity. This provides a first hint to the relation between shell effects and unitarity: starting from the non-interacting case and increasing the scattering length we see the shell effects gradually vanish. At a small scattering length, $\ac=-0.25$, shell-closures persist up to large particle numbers with small shifts due to the attractive interaction, while at the intermediate case of $\ac=-1$ only minor evidence of shell-closures can be seen and only at small particle numbers $N<20$. Finally, at the unitary limit of infinite scattering length shell effects vanish entirely.

This elimination of shell effects at unitarity is consistent with the findings of Ref.~\cite{Carlson_Gandolfi_2014} where it was shown that a simple bosonic density functional can describe unitary fermions suggesting that at their most interacting regime, fermions can acquire bosonic traits. One could argue that the elimination of a shell-structure follows from the scale-invariance expected at unitarity, but it's important to note that the harmonic trap introduces a scale that could define shell-structure and that is not seen in Fig.~\ref{fig:as}. Similarly, the shell-effects of a homogeneous system in a finite simulation box were not observed in Ref.~\cite{Forbes_Gandolfi_Gezerlis_2011}. In total, the shell structure is washed away by the strong correlations of the unitary fermions.

\begin{figure}[t!]
\includegraphics[width=0.98\linewidth]{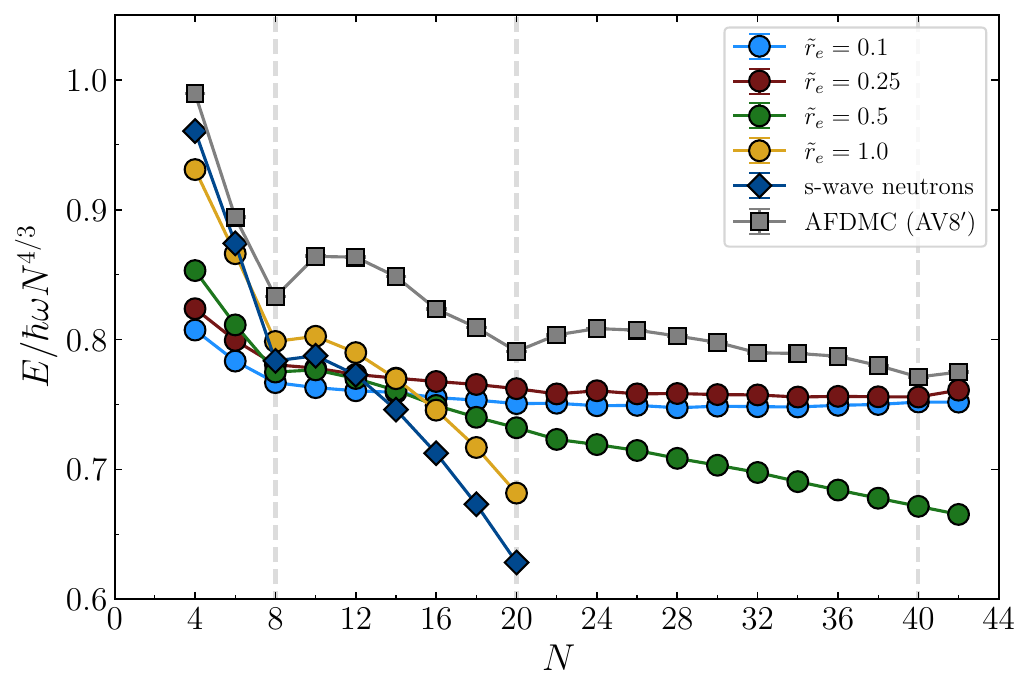}
\caption{The ground state energy (in units of $\hbar\omega N^{4/3}$) approaching unitarity in a range of effective ranges $\rec=0.1,0.25,0.5,1$ plotted with blue, red, green, and yellow circles, respectively. We plot in dark blue diamonds the ground state energy of $N$ atoms in a harmonic trap of $\hbar \omega = 10\ \text{MeV}$  and a two-particle interaction tuned to reproduce the known neutron-neutron scattering length and effective range $(r_e = 2.7\ \text{fm}$ and $a = -18.5\ \text{fm})$. We also include the AFDMC results of Ref.~\cite{Gandolfi_Carlson_Pieper_2011} for neutron matter}
\label{fig:res}
\end{figure}

\begin{figure}[t!]
\includegraphics[width=0.98\linewidth]{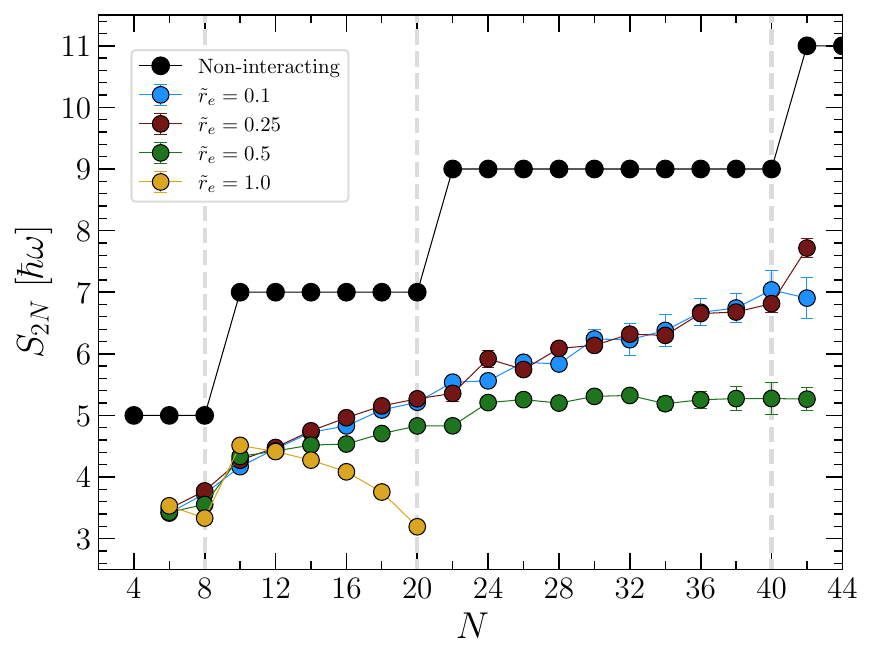}
\caption{Two particle separation energies (in units of $\hbar\omega$) approaching unitarity in a range of effective ranges $\rec=0.1,0.25,0.5,1.0$ plotted with blue, red, green, and yellow circles, respectively. To compare, we also show the two-particle separation energies in the absence of two-particle interactions where shell-effects are most pronounced.}
\label{fig:2pse}
\end{figure}

The preceding discussion invites the question of whether this behavior persists beyond unitarity and into positive scattering lengths $a>0$. In that case, the two-body interaction would be strong enough to create a free-space bound state leading to a gas of bosonic molecules. Their binding energy would cause an overall $N$-independent reduction of the energy of the system leaving $N$-dependent trends unchanged and similar to that seen at unitarity, i.e. no shell-structure.

\subsection{Effective Range}
We next turn our attention to the effects of a finite effective range when the scattering length is infinite. In Fig.~\ref{fig:res} we plot the ground state energy, scaled by the ground state energy of free fermions in a trap, for various values of the effective range $\rec=1,0.5,0.25,0.1$, that is, approaching unitarity. These variations of the effective range have a much different effect than the variations of the scattering length in Fig.~\ref{fig:as}. Starting from large effective range $\rec=1$, the system quickly binds with the inclusion of more particles, as seen by the energy dropping rapidly with $N$. This is because a large effective range comparable to the trap's length causes each particle to strongly interact with all other particles in the trap. In that case each new particle added to the system feels the attraction of all other $n=N/2$ particles in a situation comparable to a particle in an attractive mean-field. The strong binding would eventually bring the system to densities high enough to probe the repulsive effect induced by the finite effective range stabilizing the condensate. As the interaction's effective range is decreased, particles are mainly interacting with other particles in their vicinity and the binding effect goes away. At $\rec=0.5$, shell effects have decreased dramatically.  The same effect is seenin Fig.~\ref{fig:as} where, while arriving at unitarity, $\rec=0.1$,  shell effects vanish entirely. Here, the scale invariance expected at unitarity is seen once again: for large particle numbers, the ground state energy is a constant fraction of the energy of the free system, the only remaining energy scale. 

We also compare these ground state energies for various effective ranges in Fig.~\ref{fig:res} to two cases closer to that of neutrons: DMC calculations with a PT potential tuned to $a=-18.5~\textrm{fm}$ and $r_{\textrm{e}}=2.7~\textrm{fm}$, the experimental values for neutrons, and the auxiliary-field diffusion Monte Carlo (AFDMC) calculations of Ref.~\cite{Gandolfi_Carlson_Pieper_2011} which employed the AV8' potential, which provides a reasonable approximation for the nuclear interaction. In the former case, the binding effect of the finite effective range is seen again, signaled by the lsteep decrease of the energy with the $N$, while in the latter case it is not, due to the repulsive core in the $^1S_0$ channel of the neutron-neutron interaction that is present in the AV8' potential and thus prevents binding.

While shell effects should be visible in the trend of the ground state energy when increasing the particle number, e.g., Figs.~\ref{fig:as} and \ref{fig:res}, they can often be imperceptible. This is the case for the $N=20$ shell closure with $\ac$ in Fig.~\ref{fig:as}. However, they get exaggerated, and so are more easily detectable, in differential quantities such as the two particle separation energy, $S_{2N}$, and the two-particle shell gap, $\Delta_{2N}$. These are calculated as 
\begin{align}
    S_{2N}(N) &= E(N)-E(N-2)~,\\
    \Delta_{2N}(N) &= S_{2N}(N) - S_{2N}(N-2)~,
\end{align}
and can be thought of as the first and second derivatives of the ground state energy with the respect to even particle numbers. Note that, as is often the case with fermions at zero temperature, the energy of systems with even and odd particle numbers lie on two separate curves, owing to the superfluid pairing, and the distance between them, measured by the odd-even staggering (OES), is related to the pairing gap of the superfluid~\cite{Palkanoglou_Diakonos_Gezerlis_2020}. In what follows we will use the two particle separation energy and the two particle shell gap to study the shell effects present in the ground state energy. We will also calculate the pairing gap via the OES and discuss its connection to the elimination (or not) of the shell effects.

\begin{figure}[t!]
\includegraphics[width=0.98\linewidth]{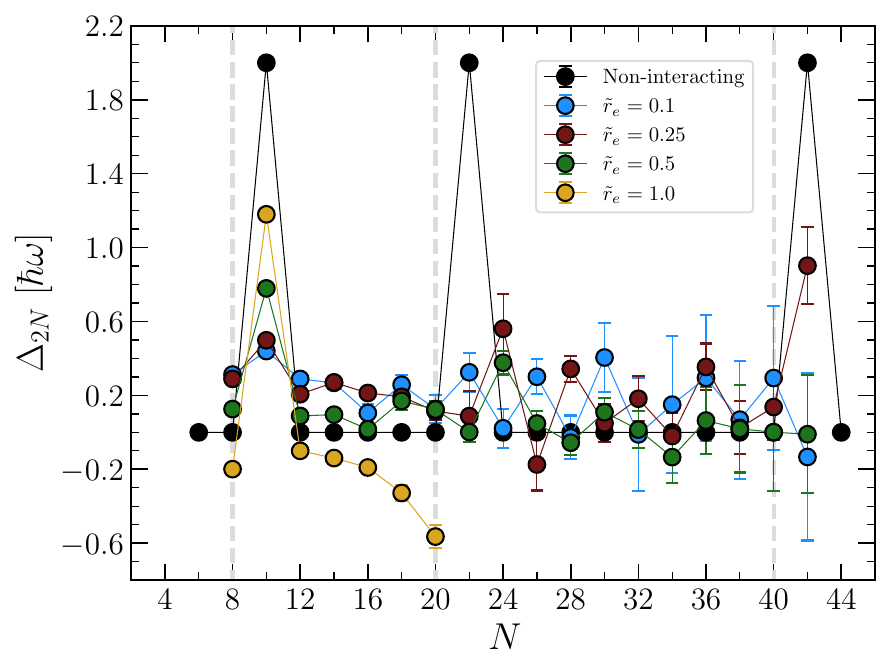}
\caption{The two particle shell gap (in units of $\hbar \omega$) approaching unitarity in a range of effective ranges $\rec=0.1,0.25,0.5,1$ plotted with blue, red, green, and yellow circles, respectively. Shell closure can be identified from pronounced peaks along the curve of fixed $\rec$. }
\label{fig:2psg}
\end{figure}

The two-particle separation energy is related to the first derivative of the energy with respect to an even number of particles, and so any shell-effects in the energy should appear as discontinuities. In Fig.~\ref{fig:2pse} we plot the two-particle separation energies for fixed infinite scattering length and a range of effective ranges, approaching the unitary limit. These calculations are compared to the non-interacting case where shell effects are expected to be most dominant. For the largest effective range considered, $\rec=1$, a shell closure is evident at $N=8$, as was the case in Fig.~\ref{fig:res}. However, the binding mechanism detailed in the discussion of Fig.~\ref{fig:res} quickly takes hold as the particle number increases and the two-particle separation decreases rapidly, signifying a deeply bound state. 
Decreasing the effective range to $\rec=0.5$, the shell-closure at $N=8$ remains present, corroborating again Fig.~\ref{fig:res}, but having avoided the strong binding, we are now able to examine higher particle numbers as well. We find a minor effect due to what could be a shifted shell-closure at $N=24$ which was largely imperceptible in Fig.~\ref{fig:res}. The next shell is expected to close at about $N=40$, but that does not occur. As we will see later, the suppression of the $N=20$ shell-closure and the vanishing of the one at $N=40$ are related to the superfluid condensation that one would expect to gradually take hold at large particle numbers. Finally, as one approaches the unitary limit, for $\rec=0.25,0.1$, the shell-effects vanish entirely, leaving a a smooth first derivative of the energy curve.

We proceed with calculating the second derivative of the energy, the two-particle shell gap $\Delta_{2N}$, plotted in Fig.~\ref{fig:2psg} for interactions approaching unitarity. For larger effective ranges, $\rec=1,0.5,0.25$, the shell closures at $N=8$ manifest as peaks in the two-particle shell gaps. In the cases of $\rec=0.5$, one more peak is seen at $N=24$ which, albeit shorter than the peak at $N=8$, stands out when compared to the otherwise relatively flat two-particle shell gap. Closer to the unitary limit, for $\rec=0.25$, a relatively weak shell effect is seen at $N=8$ followed by a relatively flat two-particle shell gap of $\sim 0.25 \hbar \omega$. The small peak at $N=24$ that was observed for $\rec=0.5$ persists here as well suggesting a shift of the $N=22$ shell closure. Finally, at $\rec=0.1$, at unitarity, $\Delta_{2N}$ does not surpass $0.5\hbar \omega$ but strongly oscillates around $0.25\hbar \omega$ for most particle numbers.

In total, using two different ways of visualizing shell effects, namely the first and second derivatives of the energy of even systems, we find that as one approaches unitarity shell effects vanish. As we will see in the next section one can see this as an effect of the strong pairing correlations in the ground state driven by the strong two-body interaction.

\subsection{Pairing Gap}
The two-particle separation energy and the two particle shell-gap calculated in the previous section provide a quantification of the shell effects. In contrast, in this section we calculate pairing gaps which quantify the pairing correlations present in the ground state. Traditionally, shell effects, and specifically shell-closures, are seen as hinting at the absence of pairing, e.g., when considering nucleons, closed-shell nuclei display vanishing pairing correlations and open-shell nuclei yield the largest pairing gaps. Conversely, and as we will try to demonstrate in this section, one can ascribe the elimination of shell-effects at unitarity to the strong pairing effects present in the ground state.

\begin{figure}[t!]
\includegraphics[width=0.9\linewidth]{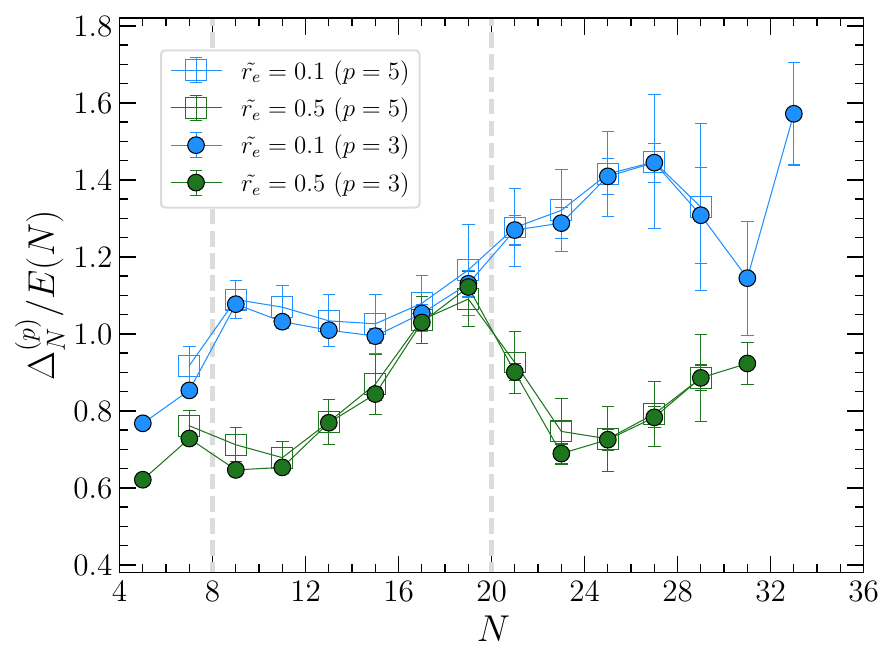}
\caption{The pairing gaps from the three-point $(p=3)$ OES approaching unitarity, for $\rec=0.1,0.5$ plotted in blue and green circles. The OES from the five-point $(p=5)$ formula is shown in dark blue and dark green squares.}
\label{fig:gap}
\end{figure}

The pairing gap is identified as the energy required to excite the ground state by breaking a pair. For finite systems the existence of the pairing gap means that the energy of even and odd systems as a function of particle number lie on two different curves separated by the pairing gap. The distance between these curves is called Odd-Even Staggering (OES) and can be extracted via finite point formulas~\cite{Palkanoglou_Diakonos_Gezerlis_2020}:
\begin{align}
    \Delta^{(3)}_N &= \frac{1}{2}\left[-E(N+1)+2E(N)-E(N-1)\right]~, \label{eq:gap3}\\
    \Delta^{(5)}_N &= \frac{1}{8}\bigg[E(N+2) -4E(N+1) +6E(N) \notag \\
    & -4E(N-1) +E(N-2)\bigg]~, \label{eq:gap5}
\end{align}
which are the $p=3-$ and $p=5-$ point formulas, respectively, where  $N$ is an odd particle number~\cite{Duguet_Bally_Tichai_2020}. The finite point formulas of Eqs.~(\ref{eq:gap3}) and (\ref{eq:gap5}) are derived assuming a slow and smooth variation with $N$ on each of the energy curves whose size is much smaller than the OES and higher $p$ finite point formulas provide better approximations for fast-varying energy curves. 
In Fig.~\ref{fig:gap} we plot the OES approaching unitarity for $p=3$ and 5. We find that both formulas agree within the statistical QMC error, therefore $\Delta^{(3)}_N$ is sufficient to describe the pairing gap $\Delta_N$. Broadly the OES rises with the particle number; this is expected because the pairing gap is an intensive quantity and the density of the trapped fermions increases with particle number. Far from unitarity, at $\rec=0.5$, the OES exhibits minima at $N=9$ and $N=23$ which are the locations of the shell closures shown in Figs.~\ref{fig:2pse} and \ref{fig:2psg}. At unitarity, i.e., $\rec=0.1$, the OES increases with no substantial minima and, on the contrary, displaying peaks at the particle numbers where shell-closures are expected.

The pronounced presence of pairing correlations in the ground states of all particle numbers at unitarity offers an explanation for the elimination of shell effects in Figs.~\ref{fig:2pse} and \ref{fig:2psg}. The strong two-particle interaction at unitarity is driving strong pairing correlations in the ground state, which in turn makes any single particle behavior unlikely. Hence, shell-effects become unlikely as well since they arise from each particle exhibiting single-particle behavior dictated by a mean-field of all other particles around it. 

\begin{figure}[t!]
\includegraphics[width=0.9\linewidth]{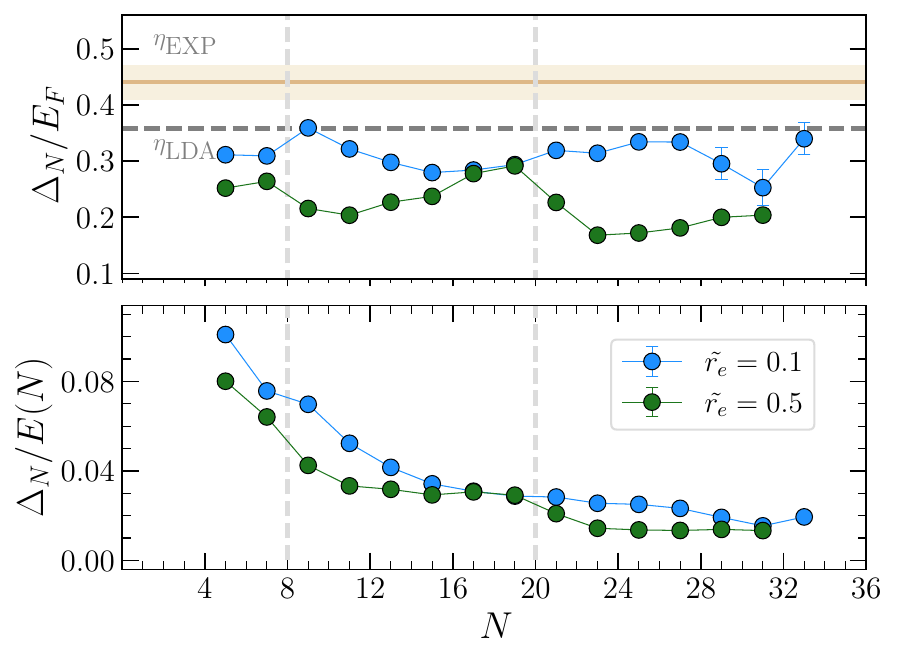}
\caption{The OES approaching unitarity, for $\rec=0.1,0.5$ plotted in blue and green circles, scaled by the Fermi energy (top panel) and the ground state energy (bottom panel). The shell-closures are displayed with dashed vertical lines. The solid lines correspond to Bertsch's parameter for the pairing gap as was measured in Ref.~\cite{Schirotzek:2008aa} and calculated for unitary Fermi gases in harmonic traps in Ref.~\cite{Haussmann:2008aa} within LDA; see text for details.}
\label{fig:gap_more}
\end{figure}

In Fig.~\ref{fig:gap_more} we scrutinize the pairing gap and its relation to the shell effects further. The upper panel shows the pairing gap scaled by the Fermi energy of trapped unitary fermions, $E_F=(3N)^{1/3}\hbar \omega$. As already discussed, in the thermodynamic limit (TL), a unitary Fermi gas is scale invariant and the pairing gap scales with the Fermi energy $\Delta_N = \eta E_F$, where $\eta$ is the Bertsch parameter for the pairing gap. In the top panel of Fig.~\ref{fig:gap_more} we include the measured value $\eta_{\textrm{EXP}}=0.44(3)$ from Ref.~\cite{Schirotzek:2008aa} as well as the calculated value $\eta_{\textrm{LDA}}=0.358$ from Ref.~\cite{Haussmann:2008aa} which was calculated within the Local Density Approximation (LDA) for a unitary Fermi gas in a harmonic trap. Our \textit{ab initio} calculations are in reasonable agreement with those within the LDA, while they differ from the measured value of $\eta_{\textrm{EXP}}$. A more detailed comparison would require the \textit{ab initio} resolution of outstanding discrepancies regarding the OES, e.g., see Ref.~\cite{Beane:2026mxi}; and we defer this task to future work. While reproducing the TL behavior of a unitary Fermi gas is not the goal of this paper and a qualitative agreement with the Refs.~\cite{Schirotzek:2008aa} and \cite{Haussmann:2008aa} is reasonable, the top panel of Fig.~\ref{fig:gap_more} also demonstrates the elimination of the shell effects as one moves closer to unitarity. Indeed the OES at $\rec=0.1$ exhibits much less variation with the particle number than it does at $\rec=0.5$. In the latter case, substantial reduction in the pairing gap is seen at the shell-closures $N=10,22$. Furthermore, at the mid-shell values of $N=18,30$, the OES becomes almost independent of $\rec$. In summary, pairing correlations grow as the interaction approaches the unitary limit and wash away the imprints of the shell-closures. This same feature is seen again but in more detail in the bottom panel of Fig.~\ref{fig:gap_more} where we show a dimensionless pairing strength: the ratio of the OES with the ground state energy, i.e., $\Delta_N/E(N)$. For $\rec=0.5$, a clear sign of the shell-structure can be seen with $\Delta_N/E(N)$ decreasing at the shell-closures. At mid-shell the $\Delta_N/E(N)$ reaches a constant plateau independent independent of $N$. For $\tilde{r}_e=0.1$, $\Delta_N/E(N)$ decreases relatively smoothly with the particle number without any signs of a shell-structure matching the $\rec=0.5$ at the mid-shell plateaus. From both panels of Fig.~\ref{fig:gap_more} a clear pattern emerges for trapped unitary gases where pairing correlations completely eliminate the shell effects as one approaches unitarity. Further, approaching unitarity, the decreasing effective range affects the pairing correlations mostly by reducing the shell-effects as the pairing gap of $\tilde{r}_e=0.1$ and $\tilde{r}_e=0.5$ are of similar size, coming apart at shell-closures at $N=10,22$ and matching at mid-shell $N=18,30$.

\section{Conclusions and Outlook}
We performed QMC calculations of the ground state energy of trapped fermions for various scattering length and effective ranges at and around unitarity. By  extracting the first and second derivatives of the energy with respect to the even particle number we studied the shell effects approaching unitarity recovering the familiar sequence of shell-closures of the harmonic trap, i.e., $N=2,8,20,40\dots$. We observed the complete elimination of shell effects at unitarity (i.e., $-1/\tilde{a}\ll 1$ and $\rec\ll 1$) for $N>4$ in all calculated quantities. 

Turning to odd particle number systems, we calculated the pairing gap via the OES. Inspecting a dimensionless pairing strength, i.e., the ratio of the OES to the ground state energy, allowed us to pin the vanishing shell effects to the strong pairing correlations present in the ground state at unitarity. The effect of the shell-structure on the pairing strength was found to be administered solely through an attenuation at the shell-closures which is otherwise washed away at unitarity. 

While these results can be probed in experiments with cold atoms, they also offer insight into neutron-rich nuclei by extension. The neutron population of a sufficiently neutron rich nucleus is expected to be dilute enough to display a behavior similar to dilute neutron matter. Furthermore, the large mismatch of the neutron and proton Fermi energies suppresses the presence of neutron-proton pairing correlations inducing to the neutron population a behavior qualitatively similar to trapped superfluid fermions. The exceptionally strong short-ranged neutron-neutron interaction would then induce a behavior similar to the unitary fermions studied here. 

In summary, we have studied the shell structure of the two-component Fermi gas in a harmonic trap at and around unitarity in search for features transferrable to the neutrons in neutron-rich nuclei. It should be noted that while a harmonic trap can support an infinite number of particles, this is not the case of the average nuclear potential which in most formulations is of finite depth resembling that of a Wood-Saxon trap. This feature adds interesting effects, such as pairing between weakly bound particles, the coupling to continuum degrees of freedom, etc. We leave such topics to future work.

\section*{Acknowledgments}
The authors would like to thank A. Gezerlis for insightful discussions.
The work of GP was supported by TRIUMF which receives federal funding via a contribution agreement with the National Research Council of Canada. 
The work of RC was supported by the Natural Sciences and Engineering Research Council (NSERC) of Canada and the Canada Foundation for Innovation (CFI). The work of S.G. was supported by the U.S. Department of Energy through Los Alamos National Laboratory (LANL). 
LANL is operated by Triad National Security, LLC, for the National Nuclear Security Administration of U.S. Department of Energy (Contract No.~89233218CNA000001).
S.G. was also supported by the U.S. Department of Energy, Office of Science, Office of Advanced Scientific Computing Research, Scientific Discovery through Advanced Computing (SciDAC) NUCLEI program.

Computational resources have been provided by the Digital Research Alliance of Canada, the Los Alamos National Laboratory Institutional Computing Program, which is supported by the U.S. Department of Energy National Nuclear Security Administration under Contract No. 89233218CNA000001, and by the National Energy Research Scientific Computing Center (NERSC), which is supported by the U.S. Department of Energy, Office of Science, under Contract No. DE-AC02-05CH11231.

\bibliography{bib}
\end{document}